\documentclass[aps,prb,superscriptaddress,floatfix,preprint,tightenlines,showpacs]{revtex4}

\usepackage{amsmath,amssymb,epsfig,color,bm}

\newcommand{\mbold}[1]{\ensuremath{\bm{#1}}}

\begin{document}

  \title{LDA+DMFT computation of the electronic spectrum of NiO}

  \author{X. Ren}
  
  \thanks{Current address: Fritz Haber Institute of the Max Planck
    Society, Faradayweg 4-6, 14195 Berlin, Germany.}

  \author{I. Leonov}
  
  \author{G. Keller}

  \author{M. Kollar}
  
  \affiliation{Theoretical Physics III, Center for Electronic
    Correlations and Magnetism, University of Augsburg, 86135
    Augsburg, Germany}

  \author{I. Nekrasov}
  
  \affiliation{Institute of Electrophysics, Ural Branch of Russian
    Academy of Science, 620016, Ekaterinburg, Russia}

  \author{D. Vollhardt}
  
  \affiliation{Theoretical Physics III, Center for Electronic
    Correlations and Magnetism, University of Augsburg, 86135
    Augsburg, Germany}
  
  \date{June 12, 2006; updated October 31, 2006}

  \begin{abstract}
    The electronic spectrum, energy gap and local magnetic moment of
    paramagnetic NiO are computed by using the local density approximation
    plus dynamical mean-field theory (LDA+DMFT). To this end the
    noninteracting Hamiltonian obtained within the local density
    approximation (LDA) is expressed in Wannier functions basis, with
    only the five anti-bonding bands with mainly Ni $3d$ character
    taken into account. Complementing it by local Coulomb interactions
    one arrives at a material-specific many-body Hamiltonian which
    is solved by DMFT together with quantum Monte-Carlo (QMC)
    simulations. The large insulating gap in NiO is
    found to be a result of the strong electronic correlations in the
    paramagnetic state.  In the vicinity of the gap region, the shape
    of the electronic spectrum calculated in this way is in good
    agreement with the experimental x-ray-photoemission and
    bremsstrahlung-isochromat-spectroscopy results of Sawatzky and
    Allen. The value of the local magnetic moment computed in the
    paramagnetic phase (PM) agrees well with that measured in the
    antiferromagnetic (AFM) phase. Our results for the electronic
    spectrum and the local magnetic moment in the PM phase are in
    accordance with the experimental finding that AFM long-range order has
    no significant influence on the electronic structure of NiO.
  \end{abstract}

  \pacs{71.27.+a, 71.30.+h, 79.60.-i}

  \maketitle

  \section{Introduction}
  \label{sec:intro}
  
  NiO is a strongly correlated electron material with a large
  insulating gap of 4.3~eV and an antiferromagnetic (AFM) ordering
  temperature of $T_{N}$=523~K.  Conventional band theories which
  stress the delocalized nature of electrons cannot explain this large
  gap and predict NiO to be metallic.\cite{Mattheiss} On the other
  hand, spin-polarized band calculations, e.g., density functional
  calculations based on the local spin-density approximation (LSDA),%
  \cite{LSDA_NiO} which do find an AFM insulating ground state in
  NiO, produce a band gap and local magnetic moment which are
  considerably smaller than the experimental values. These facts are
  often taken as evidence for the inapplicability of conventional band
  theories to strongly correlated systems like NiO.
  Indeed, already a long time ago Mott \cite{Mott} had shown that NiO
  and similar insulators may be better understood within a real-space
  picture of the solid, where localized electrons are bound to atoms
  with incompletely filled shells. This leads to the formation of
  incoherent bands, the lower and upper Hubbard bands, which are
  separated by a correlation gap of the order of the local Coulomb
  repulsion $U$. For this reason NiO has long been viewed as a
  prototype ``Mott insulator''.\cite{Mott, Brandow}
  
  This view of NiO was later replaced by that of a ``charge transfer
  insulator'',\cite{Zaanen_Sawatzky_Allen} after Fujimori and Minami
  had successfully explained the photoemission data in terms of a
  cluster model treated within the configuration-interaction method.%
  \cite{Fujimori_clusterCI} In particular, this interpretation was
  supported by the combined x-ray-photoemission (XPS) and
  Bremsstrahlung-isochromat-spectroscopy (BIS) measurements of
  Sawatzky and Allen.\cite{Sawatzky_Allen_PES}  Within this new
  picture an additional ligand $p$ band appears between the lower and
  upper Hubbard bands, and the insulating gap is formed between
  the ligand $p$ band and the upper Hubbard $d$-band. However, unless
  the $p$-$d$ hybridization is taken into account, this picture is
  still an oversimplification. Namely, the hybridization between
  transition-metal $d$ and ligand $p$ states will lead to some
  $d$-electron features also in the upper valence bands. Indeed,
  subsequent studies suggested that the first valence peak is actually
  a bound state arising from the strong hybridization of Ni $3d$ and O
  $2p$ states,\cite{ARPES_Shen91,Shen95} such that NiO is close to
  the intermediate regime of the Zaanen-Sawatzky-Allen scheme.%
  \cite{Zaanen_Sawatzky_Allen} Despite the success of the cluster
  approach it has apparent drawbacks since it neglects the band
  aspects of O $2p$ states completely which are known to play an
  important role in NiO.\cite{ARPES_Shen91,ARPES_Kuhlenbeck} The
  translational symmetry has been taken into account to some extent
  within the cluster perturbation theory recently.%
  \cite{Eder_clusterPT05}  Another extension is the treatment of a
  larger cluster (Ni$_{6}$O$_{19}$) so that nonlocal charge transfer
  excitations can be identified.\cite{Duda_PRL06}
  
  Since the cluster approach relies on adjustable parameters to fit
  the experimental spectrum it is highly desirable to obtain a
  description of the electronic structure of NiO from first
  principles. Already within L(S)DA the O $2p$ bands can be accounted
  for quite well.\cite{ARPES_Shen91}  Attempts to go beyond L(S)DA
  are based on the self-interaction-corrected density functional
  theory (SIC-DFT),\cite{Svane_SIC-DFT} the LDA+U method,%
  \cite{Anisimov_LDAU91} and the GW approximation.%
  \cite{GW_Gunnarsson,GW_Massidda97} These methods represent
  corrections of the single-particle Kohn-Sham potential in one way or
  another, and lead to substantial improvements over the L(S)DA
  results for the values of the energy gap and local moment. Within
  the SIC-DFT and LDA+U methods the occupied and unoccupied states are
  split by the Coulomb interaction $U$, whereas within LSDA this
  splitting is caused by the Stoner parameter $I$, which is typically
  one order of magnitude smaller than $U$. Therefore, compared with
  LSDA, the SIC-DFT and LDA+U methods capture more correctly the
  physics of transition-metal (TM) oxides, and improve the results for
  the energy gap and local moment significantly. The GW method goes
  one step further by calculating the self-energy to lowest order in
  the screened Coulomb interaction $W$, and the obtained band structure shows
  better agreement with angle-resolved photoemission spectra (ARPES).%
  \cite{GW_Massidda97}  Since then a large number of studies of NiO
  have been performed along these lines,\cite{Anisimov_PES93,
    Anisimov_AIM94, Hugel_LCAO97, Shick_LDAU99, Bengone_LDAU00,
    Faleev_GW04, Li_GW05} differing in the basis used and/or the
  details of the approximations. However, in both SIC-DFT and LDA+U
  the self-energy is static, i.e., does not take correlation effects
  into account adequately and thus cannot give an accurate description
  of the electronic energy spectrum of this correlated material. As to
  the GW method, its practical applications to NiO indicate that, in
  general, it also cannot explain strong correlation effects
  sufficiently. Furthermore, different implementations of the GW
  scheme lead to quite different
  results~\cite{GW_Gunnarsson,GW_Massidda97,Faleev_GW04,Li_GW05}
  regarding the value of the insulating gap and the relative positions
  of the energy bands.
  
  Thus, although considerable progress was made in the theoretical
  understanding of NiO from first principles, several important issues
  are still open: first of all, the agreement with electronic energy
  spectra is far from satisfactory. This is not really surprising
  since the self-energies employed in the previous approaches are
  either energy-independent (i.e., genuine correlations are not
  included) or are obtained from approximations whose validity is not
  entirely clear.  Secondly, since conventional first-principles
  calculations are not able to incorporate the strong local electronic
  correlations they have, with only a few exceptions, \cite{Manghi94}
  focused on the AFM ground state, and hence attributed the
  insulating gap to the existence of long-range magnetic order. It is
  well known, however, that both the band gap and the local magnetic
  moment are essentially unchanged even above the N\'{e}el temperature.%
  \cite{Brandow} Indeed, recent experiments showed that the
  long-range magnetic order has no significant influence on the
  valence band photoemission spectra \cite{Tjernberg_PES96} and the
  electron density distributions.\cite{Jauch_elecden04} These facts
  are evidence for the strongly localized nature of the electronic
  states in NiO and cannot be understood within a Slater-type, static
  mean-field theory of antiferromagnetism.\cite{Slater51} Apparently
  the AFM long-range order itself is not the driving force behind the
  opening of the insulating gap but is merely a concomitant
  phenomenon, i.e., a \textit{consequence} of the correlation-induced
  insulating gap, rather than its origin. We are thus led to the
  question whether the properties of the strongly correlated,
  paramagnetic (PM) insulating state of NiO can be calculated within a
  theoretical framework, i.e., explicitly \textit{explained}, even in
  the absence of long-range AFM order.
  
  In this work, we therefore investigate the PM insulating phase of
  NiO using the LDA+DMFT approach. This computational scheme for the
  \emph{ab initio} investigation of strongly correlated materials was
  developed during the last decade
  \cite{ldadmft_Anisimov,ldadmft_Lichtenstein,Nekrasov00} and has led
  to important insights into the physics of strongly correlated
  electron materials; for reviews see
  Refs.~\onlinecite{LDADMFT2002a,LDADMFT2002b,ldadmft_Held03,Kotliar_review,Held05}.
  The LDA+DMFT approach combines band structure theory within the
  Local Density Approximation (LDA) with many-body theory as provided
  by dynamical mean-field theory (DMFT).
  \cite{DMFT_RMP96,Kotliar_Vollhardt_PT04} Within DMFT, a lattice
  model is mapped onto an effective impurity problem embedded in a
  medium which has to be determined self-consistently,
  \cite{Georges92} e.g., by quantum Monte-Carlo (QMC) simulations.%
  \cite{Jarrell92} This mapping becomes exact in the limit of
  infinite dimensions.\cite{Metzner_Vollhardt_89}
  
  The LDA+DMFT approach is a particularly suitable scheme also for the
  investigation of the electronic properties of NiO. Namely, it takes
  into account both the material specific aspects as well as the
  strong electronic correlations in NiO, thus having advantages that
  the previous methods lack. This approach has been used by Savrasov
  and Kotliar \cite{Savrasov03} to study the phonon spectrum of NiO
  and MnO. In this work we concentrate on the electronic spectrum.
  
  The paper is organized as follows.  In section
  \ref{sec:Comput_Scheme} the LDA+DMFT computational scheme for NiO is
  presented. The general idea of modeling NiO by a material-specific
  many-body Hamiltonian is discussed in section \ref{sec:Model_Hami}.
  In section \ref{sec:LDA_band} the Wannier functions and
  single-particle Hamiltonian matrix are constructed within the LDA, and
  the correlations are included by DMFT(QMC) in section
  \ref{sec:DMFT_QMC}. The results are discussed and compared with
  experiment in section \ref{sec:result}, and finally we conclude this
  paper in section \ref{sec:conclusion}.

  \section{Computational method}
  \label{sec:Comput_Scheme}

  \subsection{Description of NiO with a multi-band Hubbard model}
  \label{sec:Model_Hami}
  
  In practice, LDA calculations usually involve a large number of
  valence $s$, $p$, and $d$ orbitals associated with all the atoms in the
  unit cell.  Since multi-orbital QMC calculations are computationally
  expensive, especially at low temperatures, not all of these orbitals
  can be included in the DMFT calculation. For this reason one needs
  to project out most of the orbitals except for the most relevant
  ones.
  
  For a large number of transition-metal compounds, the most relevant
  orbitals responsible for the physical properties are the
  transition-metal valence $d$ orbitals which are partially filled and
  are located around the Fermi level.  However, for some materials,
  e.g., the late transition metal oxides, which are either
  charge-transfer insulators or in the intermediate regime of the
  Zaanen-Sawatzky-Allen scheme, \cite{Zaanen_Sawatzky_Allen} the
  oxygen $2p$ orbitals are as important as the transition-metal
  valence $d$ orbitals, and hence have to be taken into account in the
  effective model Hamiltonian as well. Thus, in principle, the strong
  Coulomb interaction among the $d$ electrons and the $p$-$d$
  hybridization should be included and investigated on the same level.
  In DMFT this would correspond to the investigation of a
  material-specific multi-orbital version (five $d$ orbitals and three $p$
  orbitals) of the periodic Anderson model with nearest-neighbor
  hybridization. Such a comprehensive DMFT treatment is very difficult
  to perform even at present. Therefore, as a first step we turn to an
  approximate, but useful procedure: namely, we include the
  hybridization only implicitly in the construction of the set of
  Wannier functions as will be described below.\cite{wannier}
  With this set of Wannier functions as a basis, we arrive at an effective multi-band
  Hubbard model. The on-site Coulomb interaction for this Hamiltonian
  is then an effective interaction $U_{\text{eff}}$ for Wannier
  orbitals which is different from the bare Coulomb interaction $U$
  between the $d$ electrons.  In this way we are actually mimicking
  the charge transfer gap by a Mott gap.  Such a simplified treatment
  is quite analogous to the reduction of the two-band $p$-$d$
  hybridized model to a one-band Hubbard model in the context of
  cuprates.\cite{Dagotto94} (We note in passing that the Hubbard
  model and the periodic Anderson model with nearest neighbor
  hybridization display a striking similarity near the metal-insulator
  transition.\cite{Held_Bulla00}) Our simplified treatment of the
  oxygen $2p$ bands and the $p$-$d$ hybridization should be regarded
  as a first step towards a full LDA+DMFT investigation of the NiO
  problem. Within this approximation some of the spectral weight is
  lost and, in particular, the satellite structure at higher binding
  energies is ignored. Thus we are not aiming at a calculation of the
  full spectrum, but rather concentrate on the spectrum near the gap.
  
  Wannier functions can be constructed in different ways.
  Historically there were attempts \cite{Koster53, Parzen53, Kohn73}
  to produce Wannier functions by a variational procedure, without
  knowing the actual Bloch functions.  Within the context of LDA+DMFT,
  Andersen and Saha-Dasgupta \cite{Andersen00} proposed a way of
  constructing Wannier functions \textit{a priori} based on
  $N$th-order muffin-tin orbitals (NMTO) which was then applied to
  $3d^{1}$ perovskites \cite{Pavarini04} and Ti$_{2}$O$_{3}$.
  \cite{Poteryaev04} On the other hand, Wannier functions can be
  calculated straightforwardly by a unitary transformation of a set of
  Bloch bands.  However, due to the \textit{non}uniqueness of
  Wannier functions, quantitative computations along this line did not
  appear until recently. An important progress was made by Marzari and
  Vanderbilt \cite{Marzari97} who devised an optimization procedure to
  obtain the maximally localized Wannier functions. However, the
  maximal localization of Wannier functions is not a strict
  requirement if one is not interested in the orbitals themselves but
  in a proper representation of the Hamiltonian, which is the case in
  this work.

  In our paper, we follow a procedure proposed recently by Anisimov \emph{et al.},%
  \cite{wannier_paper} in which a set of $d$-like Wannier functions \cite{wannier}
  is constructed in such a way that they (i) preserve the symmetries
  of the atomic-like $d$ orbitals, and (ii) implicitly incorporate the
  admixture of oxygen $2p$ orbitals arising from the hybridization
  effects.  The LDA band structure is encoded in a non-interacting
  Hamiltonian within this basis set of Wannier functions and used as the input for
  the DMFT study. In the above-mentioned work, \cite{wannier_paper}
  calculations were carried out for SrVO$_{3}$ and V$_{2}$O$_{3}$.
  Here we extend this scheme to NiO.
  
  {}From the LDA+DMFT calculation, one can obtain the energy gap, the
  local magnetic moment, and the electronic spectrum. Since these
  quantities do not significantly depend on temperature we compare our
  results with experimental data at low temperatures and other
  theoretical results for the ground state.

  \subsection{LDA band structure and construction of Wannier functions}
  \label{sec:LDA_band}
  
\begin{figure}[tbp]
    \centerline{%
      \includegraphics[clip=true,width=0.6\textwidth]{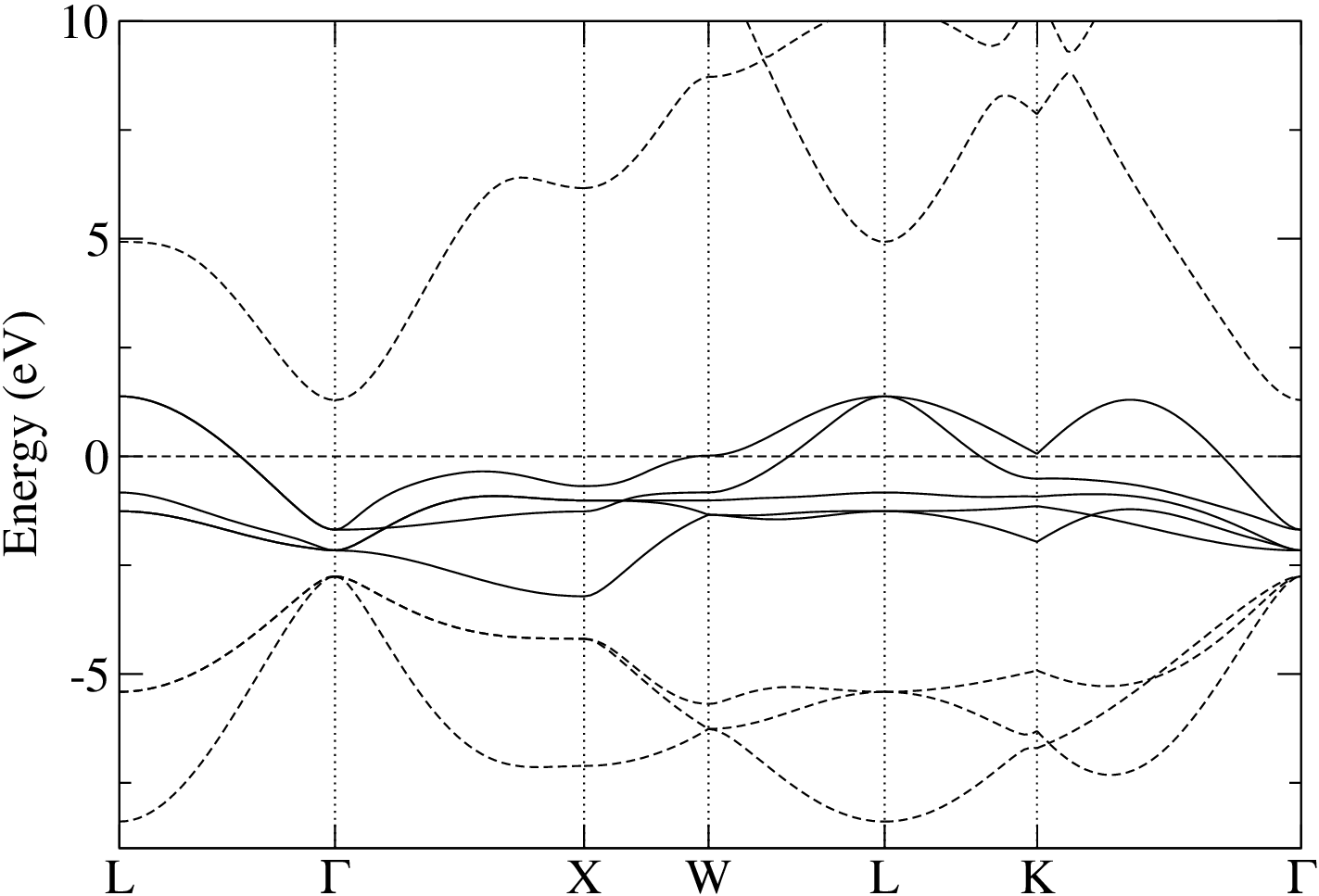}%
    }
    \caption{Nonmagnetic band structure of NiO
      obtained by the LMTO method; the Fermi energy is set to zero.
      The five ``$d$-like'' Bloch bands (in solid lines) are used for
      constructing Wannier functions.}
    \label{nio_band}
  \end{figure}

  We first perform a standard LDA band calculation using the
  linear muffin-tin orbital (LMTO) method in atomic sphere 
  approximation\cite{Andersen_LMTO75} method with the combined correction
  term included. The Stuttgart LMTO code version 47 is used for this 
  calculation. Paramagnetic NiO has rock salt crystal structure, and
  below $T_{N}$ there is a tiny rhombohedral distortion of the cubic
  unit cell. The lattice constant shows a small increase from about
  4.17~\AA{} to 4.20~\AA{} in a temperature range from 7~K to 700~K.%
  \cite{Bartel_LattConst71} Such an increase only causes a small
  deviation of the bands relatively far away from the Fermi energy,
  but no noticeable change of the Ni $3d$-dominant bands on which we
  will focus later. Since we will compare our result with the
  low-temperature experimental data, we choose the lattice constant
  $a=4.17$~\AA{} throughout this work. The calculated nonmagnetic band
  structure of NiO is shown in Fig.~\ref{nio_band} and it is in
  agreement with those published in the literature.%
  \cite{ARPES_Shen91,Eder_clusterPT05}
  
  Based on the solution of the LDA band problem, we start to construct
  a set of \textit{non}orthogonal Wannier functions centering at a lattice site \mbold{T},
  following Ref.~\onlinecite{wannier_paper},
  \begin{equation}
    |{\widetilde{W}}_{n}^{\mbold{T}}\rangle
    =
    \frac{1}{\sqrt{N}}\sum_{\mbold{k}}
    e^{i\mbold{k\cdot T}}|{\widetilde{W}}_{n\mbold{k}}\rangle
    =
    \frac{1}{\sqrt{N}}\sum_{\mbold{k}}e^{i\mbold{k\cdot T}}
    \sum_{i=N_{1}}^{N_{2}}|\psi_{i\mbold{k}}\rangle \langle\psi_{i\mbold{k}}|\phi_{n}^{\mbold{k}}\rangle
    .  \label{WF_nonorth}
  \end{equation}%
  Here $|\psi_{i\mbold{k}}\rangle$ are the single-particle Bloch
  states and $|\phi_{n}^{\mbold{k}}\rangle$ are the LMTO basis
  functions. Furthermore, $N$ is the number of discretized \mbold{k}
  points in the first Brillouin zone, and $N_{1}$ and $N_{2}$ determine
  the range of the Bloch bands to be included in the construction. In
  the case of NiO we select the five energy bands around the Fermi
  level, shown by the solid lines in Fig.~\ref{nio_band}, for our
  present calculation, and the subscript $n$ in
  $|\phi_{n}^{\mbold{k}}\rangle$ enumerates the five Ni $3d$ LMTOs.
  These five Bloch bands are dominated by Ni $3d$ states, but also
  have considerable contributions from O $2p$ states resulting from
  the hybridization effect.  The quantities
  $|{\widetilde{W}}_{n\mbold{k}}\rangle$ are simply the Fourier
  transformation of the real-space Wannier functions
  $|{\widetilde{W}}_{n}^{\mbold{T}}\rangle$, but in the multi-band
  case they should be distinguished from eigenfunctions. Hereafter, we
  refer to $|{\widetilde{W}}_{n\mbold{k}}\rangle$ as Wannier
  functions, too.
  
  So far we have a set of five \textit{non}orthogonal Wannier functions with $d$
  symmetry at every \mbold{k} point. To orthogonalize them, one needs
  to calculate the overlap matrix $O(\mbold{k})_{nn^{\prime}} =
  \langle{\widetilde{W}}_{n\mbold{k}} |
  {\widetilde{W}}_{n^{\prime}\mbold{k}}\rangle$ and its inverse square
  root. The orthogonalized Wannier functions are given in the standard way,
  \begin{equation}
    |W_{n\mbold{k}}\rangle
    =
    \sum_{n^{\prime}}
    |{\widetilde{W}}_{n^{\prime}\mbold{k}}\rangle
    (O^{-\frac{1}{2}}(\mbold{k}))_{n^{\prime}n}.
    \label{WF_orth}
  \end{equation}%
  By using Eqs.~(\ref{WF_nonorth}) and (\ref{WF_orth}) and noticing that
  $\widehat{H}|\psi_{i\mbold{k}}\rangle =
  \varepsilon_{i\mbold{k}}|\psi_{i\mbold{k}}\rangle$ and
  $\langle\phi_{n}^{\mbold{k}}|\psi_{i\mbold{k}}\rangle =
  c_{ni}^{\phantom{\ast}}(\mbold{k})$ where
  $\varepsilon_{i}(\mbold{k})$ and
  $c_{ni}^{\phantom{\ast}}(\mbold{k})$ are the eigenvalues and
  eigenvectors of the single-particle Kohn-Sham equations, one can
  express the LDA Hamiltonian matrix within the basis of
  orthogonalized Wannier functions as
  \begin{eqnarray}
    H^{\text{WF}}(\mbold{k})_{nn^{\prime}}
    &=&
    \langle W_{n\mbold{k}}|\widehat{H}|W_{n^{\prime} \mbold{k}}\rangle
    \notag \label{Ham_matr}
    \\
    &=&
    \sum_{m,m^{\prime}}(O(\mbold{k})^{-1/2})_{nm}
    \sum_{i=N_{1}}^{N_{2}}
    c_{mi}^{\phantom{\ast}}(\mbold{k})
    \varepsilon_{i}(\mbold{k})
    c_{m^{\prime}i}^{\ast}(\mbold{k})
    (O(\mbold{k})^{-1/2})_{m^{\prime}n^{\prime}}.
  \end{eqnarray}%
  {}From the above equation and the relation
  $O(\mbold{k})_{nn^{\prime}} = \sum_{i=N_{1}}^{N_{2}}
  c_{ni}^{\phantom{\ast}}(\mbold{k})
  c_{n^{\prime}i}^{\ast}(\mbold{k})$, it is easy to see that the
  matrix elements of $H^{\text{WF}}(\mbold{k})$ can be built up
  exclusively in terms of the eigenvalues $\varepsilon_{i}(\mbold{k})$
  and eigenvectors $c_{ni}^{\phantom{\ast}}(\mbold{k})$ of the LDA
  single-particle problem. Details of the procedure can be found in
  Ref.~\onlinecite{wannier_paper}.
  
\begin{figure}[tbp]
    \centerline{%
      \includegraphics[clip=true,width=0.6\textwidth]{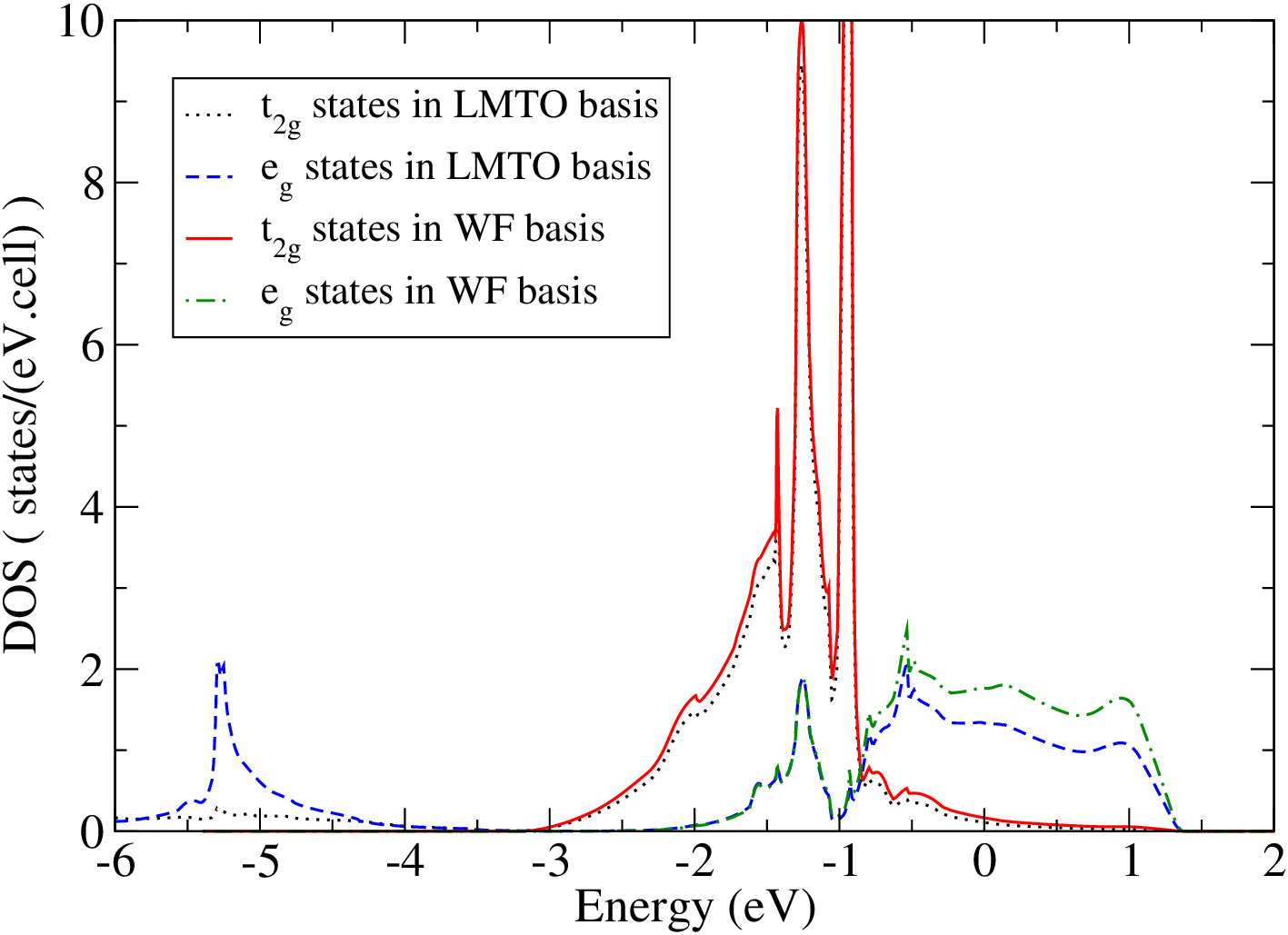}%
    }
    \caption{(color online). t$_\text{2g}$ and e$_\text{g}$ resolved LDA DOSs in the basis
      of $d$-like Wannier functions (t$_\text{2g}$-red solid curve,
      e$_\text{g}$-green dot-dashed curve) and Ni $3d$ LMTOs
      (t$_\text{2g}$-black dotted dash curve, e$_\text{g}$-blue dash
      curve).}
    \label{ldados}
  \end{figure}

  Next we present the density of states (DOS) of these Wannier functions in
  Fig.~\ref{ldados} and compare with the Ni $3d$ LMTO DOSs.
  The Wannier functions constructed in the above procedure have $d$ symmetry, and
  therefore in a cubic system as NiO they can also be classified into
  threefold degenerate $t_\text{2g}$ states and twofold degenerate
  $e_\text{g}$ states. However, one should keep in mind that these Wannier functions
  are not pure $d$ states, but rather the anti-bonding-like states
  resulting from the hybridization between Ni $3d$ and O $2p$ states.
  This is obvious from Fig.~\ref{ldados}, where in the Ni
  $3d$-dominant region the Wannier functions have more spectral weight than LMTOs,
  the extra spectral weight being due to the contribution of O $2p$
  states.  By construction, the bands at higher binding energy with
  mainly oxygen character are not included in the Wannier functions. Thus, as
  discussed above, the present scheme does not fully include the
  O $2p$ bands and the $p$-$d$ hybridization.

  \subsection{Inclusion of correlation effects by LDA+DMFT(QMC)}
  \label{sec:DMFT_QMC}
  
  The single-particle Hamiltonian matrix $H^{\text{WF}}(\mbold{k})$ in
  Wannier function basis must be supplemented with the multi-orbital Coulomb
  interaction including the local Hund's rule exchange interaction. In
  principle, $H^{\text{WF}}(\mbold{k})$ should be corrected for the
  Coulomb interaction that has already been taken into account at the
  LDA level, yielding a double-counting-free single-particle
  Hamiltonian matrix $H_{0}^{\text{WF}}(\mbold{k})$. But in the
  present case, this effect can be approximated by a trivial shift of
  the energy (for details see Ref.~\onlinecite{ldadmft_Held03}).  The full
  Hamiltonian thus reads
  \begin{eqnarray}
    H &=&
    \sum_{\mbold{k},n,n^{\prime},\sigma}
    H_{0}^{\text{WF}}(\mbold{k})_{n,n^{\prime}}
    d_{\mbold{k}n\sigma}^{\dagger}
    d_{\mbold{k}n^{\prime}\sigma}^{\phantom{\dagger}}
    +\frac{1}{2}
    \mathop{\sum{^\prime}}_{i,n,n^{\prime},\sigma,\sigma^{\prime}}
    U_{nn^{\prime}}^{\sigma {\sigma}^{\prime}}
    d_{in\sigma}^{\dagger}
    d_{in\sigma}^{\phantom{\dagger}}
    d_{in^{\prime}{\sigma}^{\prime}}^{\dagger}
    d_{in^{\prime}{\sigma}^{\prime}}^{\phantom{\dagger}}
    \notag  \label{full_hamilt}\\
    &-&
    \frac{1}{2}\mathop{\sum{^\prime}}_{i,n,n^{\prime},\sigma}
    J_{nn^{\prime}}
    d_{in\sigma}^{\dagger}
    d_{in^{\prime}\bar{\sigma}}^{\dagger}
    d_{in^{\prime}\sigma}^{\phantom{\dagger}}
    d_{in\bar{\sigma}}^{\phantom{\dagger}}.
  \end{eqnarray}%
  The general interaction $U_{nn^{\prime}}^{\sigma \sigma^{\prime }}$
  denotes the Coulomb interaction matrix among different Wannier
  orbitals. For a cubic system it satisfies the relation
  $U_{nn^{\prime}}^{\sigma
    \sigma^{\prime}}=U-2J(1-\delta_{nn^{\prime}})-J\delta_{\sigma
    \sigma^{\prime}}$ to a good approximation. Here $U$ corresponds to
  the \textit{effective} interaction between electrons in Wannier
  orbitals as discussed above and not to the \textit{bare} interaction
  between the actual $d$ electrons. The values of the parameters $U$
  and $J$ are accessible from a constrained LDA calculation.
  Specifically, from a constrained LDA calculation the average Coulomb
  interaction $\bar{U}$, and the Hund's rule exchange coupling $J$ is
  obtained, which satisfies, \cite{ldadmft_Held03}
  \begin{equation}
    \bar{U}=\frac{U+(M-1)(U-2J)+(M-1)(U-3J)}{2M-1},
    \label{Ubar}
  \end{equation}%
  where $M=5$ is the number of the interacting orbitals. Given
  $\bar{U}$ and $J$ one can thus obtain $U$ from Eq.~(\ref{Ubar}).  Since
  the constrained LDA calculation has not been performed for the Wannier functions,
  we take the $\bar{U}$ value for the LMTO basis from the
  literature,\cite{Anisimov_LDAU91}
  and estimate the value for the Wannier function basis by
  $\bar{U}^{\text{WF}}=\bar{U}^{\text{LMTO}}(1-x)^{2}$ where $x$ is
  the admixture of oxygen $2p$ states into the
  Wannier functions.\cite{Anisimov_priv_comm} For NiO, $x$ is approximately $0.15$,
  and $\bar{U}^{\text{LMTO}}=8$~eV.\cite{Anisimov_LDAU91} This means
  that $\bar{U}^{\text{WF}}\ \approx 5.8$~eV. Concerning the $J$ value
  it is reasonable to assume that it is only weakly basis-dependent,
  and we just take the LMTO value $J=1$~eV here. With these
  considerations we obtain $U=8$~eV for the Wannier function basis.
  
  We solve the above material-specific model (\ref{full_hamilt}) by
  means of DMFT, \cite{Kotliar_Vollhardt_PT04, DMFT_RMP96,
    DMFT_Vollhardt_JPSJ05} which maps the original lattice model onto
  a single-impurity model subject to a self-consistent condition. The
  impurity problem is in turn solved by quantum Monte-Carlo (QMC)
  techniques, \cite{Hirsch_QMC86} yielding an imaginary-time Green
  function. The maximum-entropy method (MEM) \cite{Jarrel_MEM96} is
  then used to obtain the physical spectral function on the real
  frequency axis. We neglect the spin-flip term in Eq.~(\ref{full_hamilt})
  which is known to lead to ``minus-sign'' problems in QMC simulations,%
  \cite{ldadmft_Held03} assuming that this term does not play a
  significant role here. QMC simulations can only be performed at not
  too low temperatures, since the computation effort scales with
  $1/T^{3}$.

  \section{Results and comparison with experiment}
  \label{sec:result}
  
\begin{figure}[tbp]
    \centerline{%
      \includegraphics[clip=true,width=0.6\textwidth]{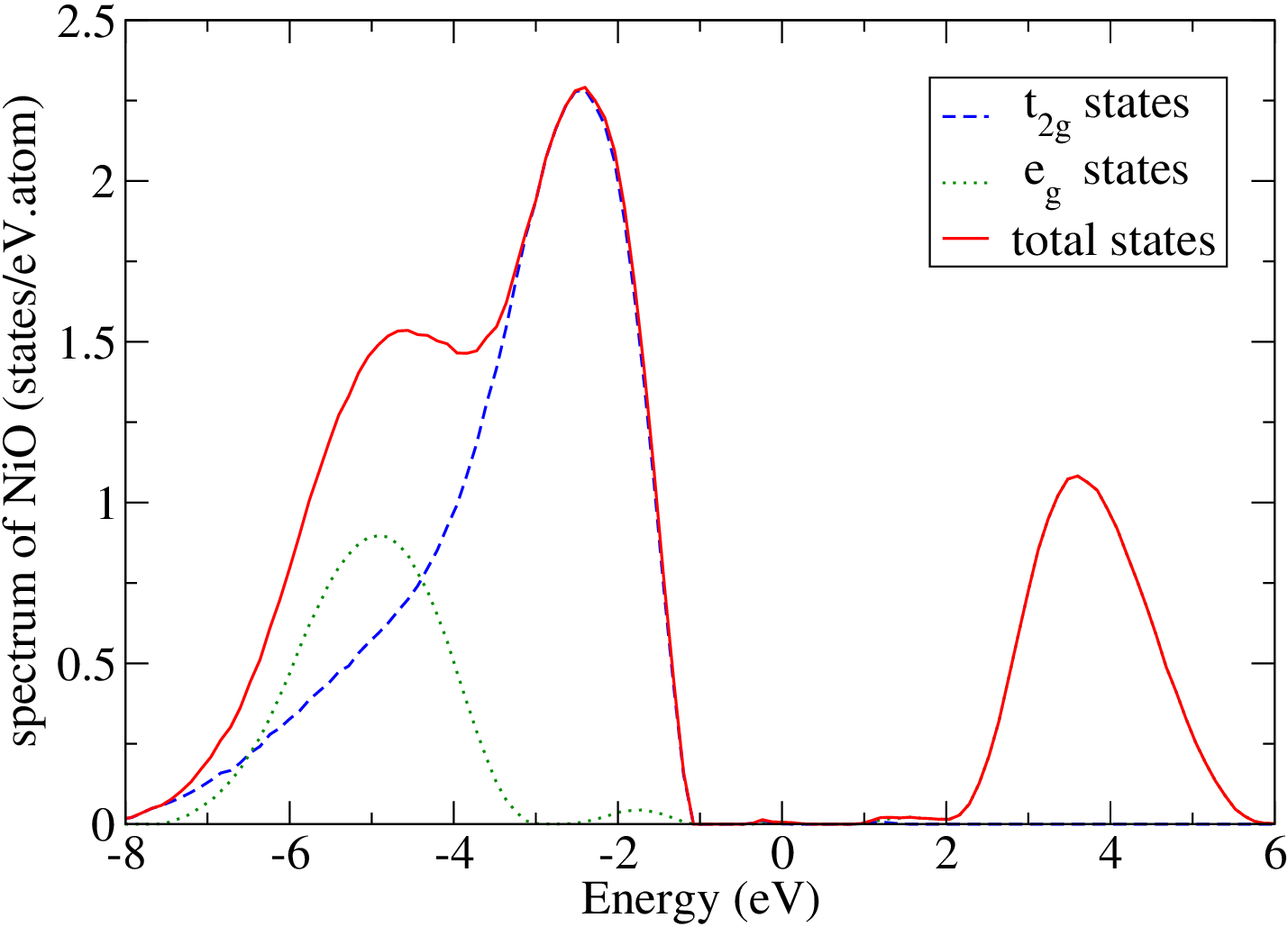}%
    }
    \caption{(color online). Theoretical energy spectrum of NiO obtained by the
      LDA+DMFT approach for $T=1160$~K, and $U=8$~eV, $J=1$~eV.
      t$_{2g}$ states (blue dashed curve) and e$_{g}$ states (green
      dotted curve) are resolved, and total states (red solid curve)
      are the sum of the two. Note the curve for total states
      coincides with that of the unoccupied e$_{g}$ states above the
      chemical potential.}
    \label{DMFT_u8j1_1160}
  \end{figure}

  Our DMFT(QMC) calculations were performed at temperatures $T=1160$~K
  and $725$~K respectively. In QMC simulations we used 40 time slices
  for the former case and 64 for the latter, and in both cases up to
  $10^6$ QMC sweeps were made to ensure good convergence.
  The electronic energy spectrum obtained at temperature $T=1160$~K
  for interaction parameters $U=8$~eV and $J=1$~eV is shown in
  Fig.~\ref{DMFT_u8j1_1160}.
  The dominant peak at about $-2.5$~eV is
  due to the t$_\text{2g}$ bands which are hence completely filled.
  The e$_\text{g}$ bands are split into lower and upper Hubbard bands.
  The insulating gap is therefore situated between the occupied
  t$_\text{2g}$ bands and unoccupied e$_\text{g}$ bands.\cite{MEM}
  The shoulder between $-6$~eV and $-4$~eV originates from the
  occupied e$_\text{g}$ bands. We understand that this picture differs
  from the one provided by the cluster model,%
  \cite{Fujimori_clusterCI} where both the main peak and side peak
  are dominated by oxygen character.  Indeed, in the present
  calculation, the hybridization between Ni $3d$ states and O $2p$
  states is fixed at the LDA level, and a more complete theory of NiO
  should also allow the $p$-$d$ hybridization to evolve under the
  influence of the interaction among $d$ electrons. As discussed
  earlier this requires an explicit inclusion of oxygen states which
  is not implemented in the present work.  Keeping the drawback of the
  present scheme in mind, we will leave the resolution of this problem
  to future studies, and take the present result as a first
  approximation.  But as shown below, the agreement between the energy
  spectrum obtained in this calculation and the experiment is already
  very good.
 
\begin{figure}[tbp]
    \centerline{%
      \includegraphics[clip=true,width=0.6\textwidth]{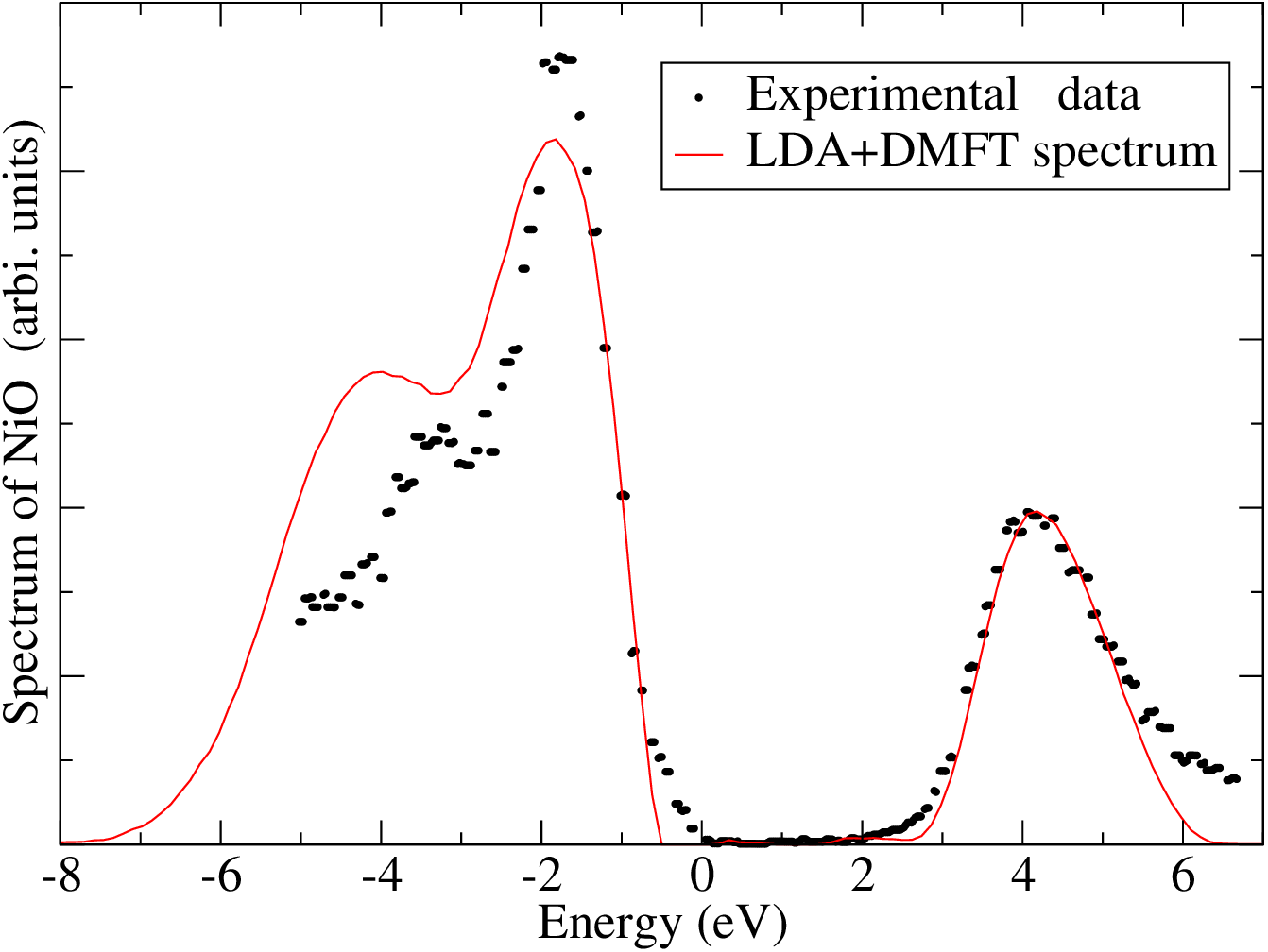}%
    }
    \caption{(color online). Theoretical spectrum is obtained by the LDA+DMFT
      approach (red curve) for $U=8$~eV and $J=1$~eV at $T=1160$~K,
      compared with experimental XPS+BIS data (black dots) after
      Sawatzky and Allen.\protect\cite{Sawatzky_Allen_PES} The zero
      energy point of the theoretical curve is shifted to fit the
      energy scale of the experimental data.}
    \label{NiO_PES}
  \end{figure}

  In Fig.~\ref{NiO_PES} our LDA+DMFT results are compared with data of
  the combined XPS-BIS experiment by Sawatzky and Allen%
  \cite{Sawatzky_Allen_PES} for the low-temperature AFM phase.
  Although our calculation was performed for the high-temperature PM
  phase a comparison is meaningful since the electronic structure of
  NiO is almost unaffected by the magnetic phase transition.%
  \cite{Tjernberg_PES96,Jauch_elecden04} In view of the ambiguity in
  determining the position of the chemical potential in the insulating
  gap we shifted the spectrum appropriately with respect to the
  experimental data.  The overall agreement between the calculated
  single-particle spectrum and the experimental data is surprisingly
  good. Only the shoulder below the main peak is not very well
  reproduced. It should be noted that the energy spectrum in this
  region, i.e., far away from the Fermi level, is difficult to obtain
  accurately due to the limitations imposed by MEM.  On the other
  hand, matrix element effects, which have so far received little
  attention in the theoretical calculation of the XPS of NiO, may also
  play a role here.  Although detailed quantitative information on
  matrix element effects is not available for NiO, it is generally
  known that in the XPS regime (i) O $2p$ features become weak
  relative to the Ni $3d$ features,\cite{Eastman_PRL75} and (ii) the
  photoionization cross section of transition metals is suppressed at
  higher binding energies compared to lower ones.
  \cite{Nahm93,Speier88} The first effect is unlikely to be
  significant since oxygen features are only contained indirectly in
  the calculation. The second effect may even improve the agreement
  further by reducing the weight of the theoretical spectrum at higher
  binding energies. As a whole, we do not expect matrix element
  effects to impair the agreement of our results with experiment.
  
\begin{figure}[tbp]
    \centerline{%
      \includegraphics[clip=true,width=0.6\textwidth]{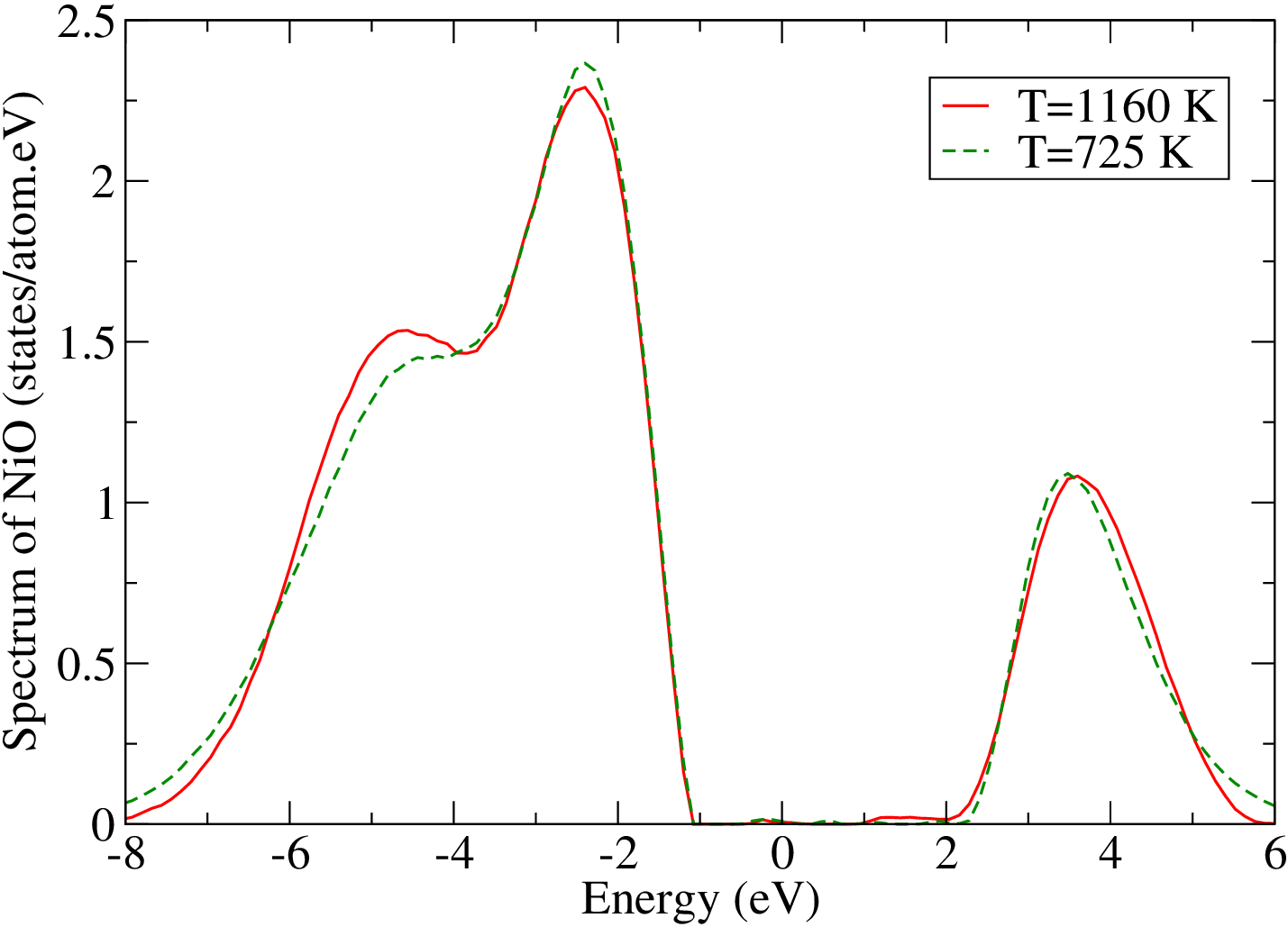}%
    }
    \caption{(color online). Theoretical spectrum of NiO obtained by the
      LDA+DMFT approach for $U=8$~eV, $J=1$~eV at $T=1160$~K (red
      solid curve) and $T=725$~K (green dashed curve) respectively.}
    \label{DMFT_temp_comp}
  \end{figure}

  To analyze the effect of the QMC simulation temperature on the
  spectrum we also performed calculations at $T=725$~K. The result is
  shown in Fig.~\ref{DMFT_temp_comp} in comparison with that obtained
  at $T=1160$~K.
  Within the numerical error of QMC and MEM the
  spectra at those two temperatures do not show any significant
  difference.  For this reason we expect our results to describe the
  spectrum of paramagnetic NiO also at even lower temperatures.
 
\begin{figure}[tbp]
    \centerline{%
      \includegraphics[clip=true,width=0.6\textwidth]{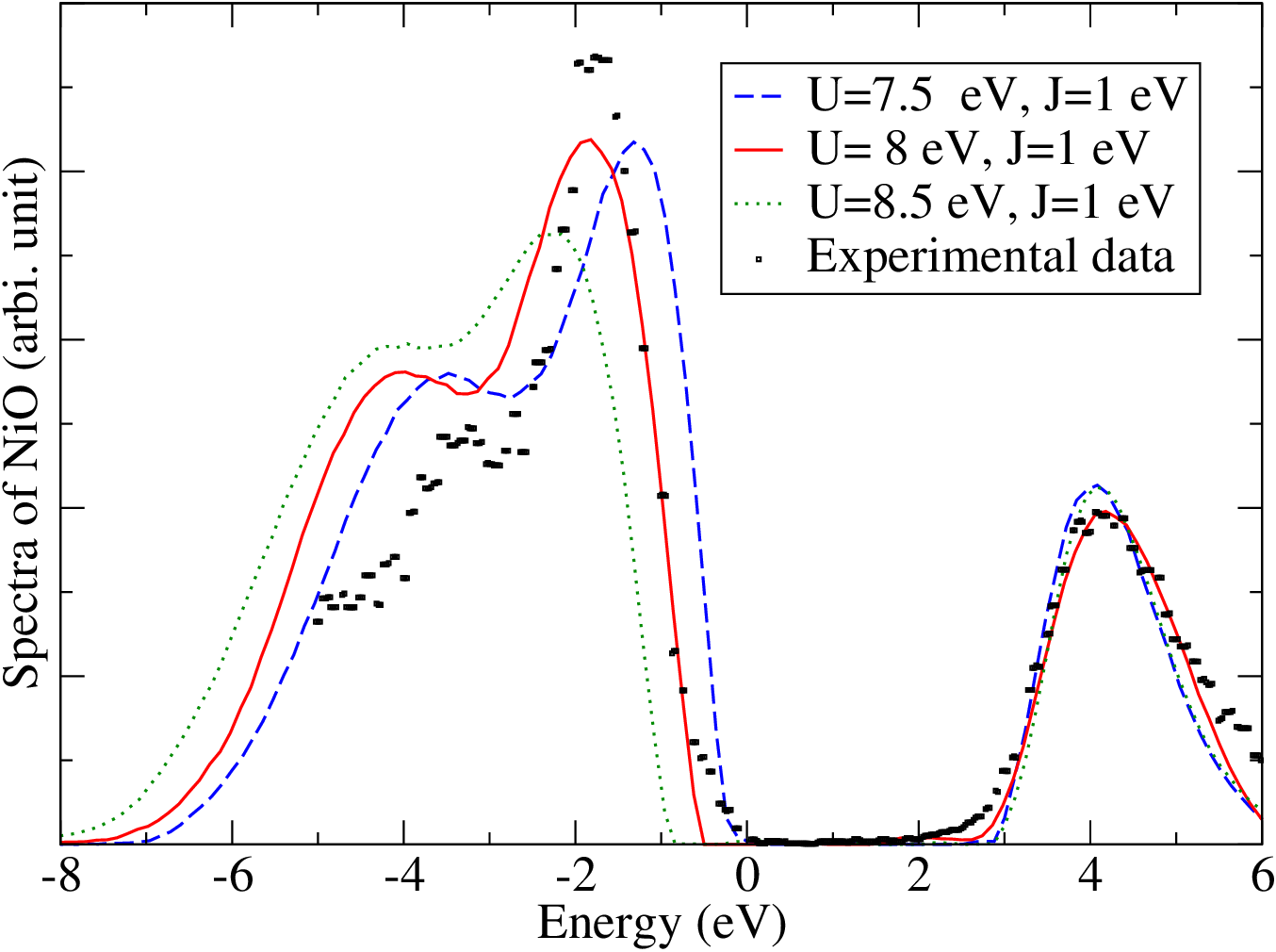}%
    }
    \caption{Theoretical spectra obtained by the LDA+DMFT 
      approach for $U=7.5$, $8$, and $8.5$~eV (blue dashed, red solid,
      and green dashed curves) respectively with fixed $J=1$~eV and
      temperature $T=1160$~K.  Experimental XPS+BIS data from Sawatzky
      and Allen \protect\cite{Sawatzky_Allen_PES} are also shown for
      comparison.}
    \label{DMFT_U_comp}
  \end{figure}

  It is also interesting to investigate how the spectrum changes when
  the interaction parameter $U$ is changed. For this purpose we
  performed the calculations also for $U=7.5$~eV and $8.5$~eV, with
  fixed $J=1$~eV, at $T=1160$~K. The results are shown in
  Fig.~\ref{DMFT_U_comp}, in comparison with that for $U=8$~eV and the
  experimental data.
  Here, to illustrate the change of the insulating
  gap for different $U$ values, the positions of the conduction-band
  peaks were kept fixed. {}From Fig.~\ref{DMFT_U_comp} we see that the
  best agreement with experiment is indeed found for $U=8$~eV --- the
  value determined from the constrained LDA calculation. The Mott gap
  is seen to increase roughly linearly with $U$. The t$_\text{2g}$ and
  e$_\text{g}$ splitting, however, does not change since it is
  controlled by the energy $J$ whose value is fixed here.

  In Table~\ref{Gap_Mom} we present our results for the energy gap and
  the local magnetic moment in comparison with experimental results
  and the results of other theoretical approaches.
  Here we measured
  the energy gap of 4.3~eV as the distance between the half-maximum of
  the highest valence peak and the conduction peak, respectively, as
  was done in Ref.~\onlinecite{Sawatzky_Allen_PES}. The local magnetic
  moment of 1.70 $\mu_{B}$ is obtained in an indirect way. In
  experiment the local moment is associated with the spin polarization
  in the immediate vicinity of the Ni ion. However, in the DMFT
  calculation the magnetic moment is associated with the Wannier functions.  For the
  parameter $U=8$~eV and $J=1$~eV, we find a magnetic moment
  associated with the Wannier functions as $m\approx 1.99~\mu_B $.\cite{Local_Mom}
  We can then estimate the local magnetic moment associated with the
  Ni ion by the relation $m_{\text{loc}}=m(1-x)$ where $x=0.15$ is
  again the contribution of O $2p$ states to
  Wannier functions.\cite{Anisimov_priv_comm}  This leads to $m_{\text{loc}}\approx
  1.70~\mu_B$ which is in accordance with the measurement and theoretical
  results obtained for the AFM phase (see Table \ref{Gap_Mom}).
  Apparently the value of the local magnetic moment is almost
  unaffected by the AFM-PM phase transition.
  \begin{table}[b]
    \caption{The LSDA, GW, LDA+U, LDA+DMFT (present work) and experimental
      energy gaps and magnetic moments for NiO.}
    \label{Gap_Mom}%
    \begin{ruledtabular}
      \begin{tabular*}{\textwidth}{@{}%
          l@{\extracolsep{\fill}}%
          c@{\extracolsep{\fill}}%
          c@{\extracolsep{\fill}}%
          c@{\extracolsep{\fill}}%
          c@{\extracolsep{\fill}}%
          c@{}}
        & LSDA 
        & GW$^{a}$
        & LDA+U$^{b}$
        & LDA+DMFT
        & Expt.\\ \hline
        Energy gaps (eV)
        & 0.3 & 3.7 & 3.7 & 4.3 &
        4.3$^{c}$,
        4.0$^{d}$
        \\
        Magnetic moments ($\mu_B$)
        & 1.09 & 1.83 & 1.70 & 1.70 & 
        1.64$^{e}$,
        1.77$^{f}$,
        1.90$^{g}$
      \end{tabular*}
    \end{ruledtabular}
    \begin{tabular*}{\textwidth}{@{}l@{\extracolsep{\fill}}l@{}}
      $^{a}${S. Massidda \textit{et al.} \cite{GW_Massidda97}}
      &
      $^{b}${V.~I. Anisimov \textit{et al.} \cite{Anisimov_PES93}}
      \\
      $^{c}${G.~A. Sawatzky and J.~W. Allen \cite{Sawatzky_Allen_PES}}
      &
      $^{d}${S. H\"ufner \textit{et al.} \cite{Hufner_PES84}}
      \\
      $^{e}${H.~A. Alperin \cite{Alperin_EXP62}}
      &
      $^{f}${B.~E.~F. Fender \textit{et al.} \cite{Fender_EXP68}}
      \\
      $^{g}${A.~K. Cheetham and D.~A.~O. Hope \cite{Cheetham_EXP83}}
    \end{tabular*}
  \end{table}

  \section{Conclusion}
  \label{sec:conclusion}
  
  In this paper we used LDA+DMFT to compute the electronic
  single-particle spectrum of NiO. Specifically, Wannier functions
  were constructed corresponding to the five Bloch bands across the
  Fermi level from the LDA band structure. These were used as the
  basis for a material-specific multi-band Hubbard-like Hamiltonian.
  This Hamiltonian was then solved by DMFT using QMC as the impurity
  solver. The electronic spectrum obtained in this way is in very good
  overall agreement with the experimental XPS+BIS
  results~\cite{Sawatzky_Allen_PES} around the gap region. The energy
  gap and local magnetic moment obtained by us is in good agreement
  with experimental results. Taken together with the experimental
  observation that the long-range AFM order has no significant
  influence on the valence band photoemission spectra,%
  \cite{Tjernberg_PES96} our result proves that the large insulating
  gap in NiO is due to strong electronic correlations in the
  paramagnetic phase. The magnetic order in NiO is therefore only a
  secondary effect, i.e., a consequence rather than the origin of the
  gap. Hence we expect the conduction band photoemission spectra to
  remain almost unchanged by the AFM long-range order.
  
  In the construction of the \textit{ab initio} Hamiltonian
  (\ref{full_hamilt}), only the five ``anti-bonding'' bands (which have
  mainly Ni $3d$ characters in the LDA calculation) were included,
  while the three ``bonding'' bands (which are the mixture of the Ni
  $3d$ and O $2p$ states, but have more O contributions) below them
  were neglected. Therefore in the present work the contributions to
  the Wannier functions from Ni $3d$ and O $2p$ states depend on the
  LDA results. This means that the valence bands close to the Fermi
  level have mainly Ni $3d$ characters. The insulating gap we obtained
  is therefore an effective Mott-Hubbard gap between Wannier states.
  In other words, we used an effective Mott-Hubbard gap to mimic a
  charge-transfer gap \cite{Fujimori_clusterCI,Sawatzky_Allen_PES} or
  a gap with mixed character.\cite{Bengone_LDAU00,Schuler05} 
  Strictly speaking, the gap here is also of mixed character
  since some amount of oxygen contribution is contained in the Wannier 
  functions.
  The results obtained by such a treatment are surprisingly good.  
  This may be due to the fact that correlation effects are treated better
  within DMFT than within any other theoretical approach available so
  far, and also because features due to oxygen are rather suppressed
  in XPS.\cite{Eastman_PRL75}
  
  In spite of the limitations of the current implementation, we showed
  that the LDA+DMFT approach which combines first-principles,
  material-specific information with strong correlation effects is
  able to deal with late transition-metal monoxides like NiO. A more
  complete treatment of NiO within the LDA+DMFT approach will require
  the inclusion of oxygen bands and the $p$-$d$ hybridization. Only in
  this way can one produce a full spectrum of NiO and identify the
  satellite structure at high binding energies.

  \acknowledgments
  
  We are grateful to V.~I. Anisimov, K. Held, V. Eyert and J. Kunes
  for valuable discussions. This work was supported by the Deutsche
  Forschungsgemeinschaft through Sonderforschungsbereich 484, the
  Russian Basic Research Foundation grants 05-02-16301, 05-02-17244,
  the RAS Programs ``Quantum macrophysics'' and ``Strongly correlated
  electrons in semiconductors, metals, superconductors and magnetic
  materials'', Dynasty Foundation, the Grant of the President of
  Russia MK-2118.2005.02, and the interdisciplinary grant UB-SB RAS.
  Computations were performed at the John von Neumann Institut f\"ur
  Computing, J\"ulich.


\begin{thebibliography}{99}
\bibitem{MEM}%
    The finite but very low spectral weight in the gap between 1 and 2~eV appears to be an artifact of the maximum-entropy method.
\bibitem{Local_Mom}%
    To obtain a local magnetic moment in the PM phase whose value is close to the maximal value is a nontrivial result. This value becomes smaller as $U$ and $J$ are reduced.
\end{thebibliography}
\end{document}